\documentclass[prl,twocolumn,lettersize,oneside]{revtex4}
\usepackage{graphicx}
\usepackage{latexsym}
\usepackage{amsmath}
\usepackage{graphics}
\usepackage{amssymb}
\usepackage{layout}
\usepackage{verbatim}
\usepackage{amsfonts,epsfig}
\usepackage{slashed}
\usepackage{feynmp}
\usepackage{float}
\usepackage{epsfig}
\usepackage{xcolor}
\usepackage{epstopdf}
\newcommand{\ket}[1]{\left|#1\right>}

\newcommand{\beq}{\begin{equation}}
\newcommand{\eeq}{\end{equation}}
\newcommand{\bea}{\begin{eqnarray}}
\newcommand{\eea}{\end{eqnarray}}

\begin{document}

\title{Noise-induced collective quantum state preservation in spin qubit arrays}
\author{Edwin Barnes$^{1,2,3}$, Dong-Ling Deng$^{1,2}$, Robert E. Throckmorton$^{1,2}$, Yang-Le Wu$^{1,2}$ and S.~Das Sarma$^{1,2}$}
\affiliation{$^{1}$Condensed Matter Theory Center, Department of Physics, University of Maryland, College Park, Maryland 20742-4111, USA\\
$^{2}$Joint Quantum Institute, Department of Physics, University of Maryland, College Park, Maryland 20742-4111, USA\\
$^{3}$Department of Physics, Virginia Tech, Blacksburg, Virginia 24061, USA}

\begin{abstract}
The hyperfine interaction with nuclear spins (or, Overhauser noise) has long been viewed as a leading source of decoherence in individual quantum dot spin qubits. Here we show that in a coupled multi-qubit system consisting of as few as four spins, interactions with nuclear spins can have the {\it opposite} effect where they instead preserve the collective quantum state of the system. This noise-induced state preservation can be realized in a linear spin qubit array using current technological capabilities. Our proposal requires no control over the Overhauser fields in the array; only experimental control over the average interqubit coupling between nearest neighbors is needed, and this is readily achieved by tuning gate voltages. Our results illustrate how the role of the environment can transform from harmful to helpful in the progression from single-qubit to multi-qubit quantum systems.
\end{abstract}

\maketitle

Electron spins trapped in semiconductor quantum dots are natural qubit candidates due to their long coherence times and easy controllability \cite{Hanson_RMP07}. While individual spins (referred to as Loss-DiVincenzo (LD) qubits \cite{Loss_PRA98}) can be controlled directly with time-dependent magnetic fields \cite{Koppens_Nature06,Laucht_SA15} or indirectly with ac electric fields \cite{Nowack_Science07,Yoneda_PRL14}, fast dc electrical control can also be implemented in gate-defined quantum dots by encoding individual qubits in the collective spin states of two or three electrons. Singlet-triplet (ST) qubits \cite{Levy.02,Petta_Science05,Barthel_PRL09,Foletti_NP09,Bluhm_NP11,Maune_Nature12}, for example, are encoded in the singlet and unpolarized triplet states of two electrons confined in a double quantum dot, allowing them to be controlled electrically by tuning the inter-spin exchange coupling with gate voltages. These qubits require a magnetic field gradient across the double quantum dot for universal single qubit operations; this can come from a micromagnet \cite{Wu_PNAS14,Yoneda_APE15} or from nuclear spin programming \cite{Foletti_NP09,Bluhm_PRL10}. The field gradient can be avoided by creating an exchange-only (EO) qubit \cite{DiVincenzo_Nature00,Gaudreau_PRL06,Laird_PRB10,Medford_NatNano13,Medford_PRL13} from the spins of three electrons, enabling very fast purely electrical control, albeit at the expense of introducing leakage channels within the large, three-spin Hilbert space. Each approach has merits and is being actively pursued experimentally.

Despite fast control and long coherence times, decoherence continues to be a primary challenge in scaling spin qubits up to larger, multi-qubit systems. All spin qubits are subject to noise due to charge fluctuations \cite{Dial_PRL13}, which may be intrinsic to the semiconductor sample or arise in the gate electrodes used to define and control the quantum dots, and to hyperfine interactions with lattice nuclear spins (i.e., Overhauser noise) \cite{Medford_PRL12,Fink_PRL13}. Considerable work has been devoted to finding ways to mitigate both types of noise, including the improvement of samples and fabrication methods and also the design of control methods such as dynamical decoupling \cite{Uhrig_PRL07,Witzel_PRL07,Bluhm_NP11} and dynamically corrected gates \cite{Wang_NatComm12,Kestner_PRL13,Wang_PRB14,Barnes_SciRep15}. In fact, eliminating decoherence arising from Overhauser (and charge) noise is one of the central themes of spin qubit research because it is universally accepted that such noise is detrimental to preserving the qubit quantum state.

In this work, we show how Overhauser disorder due to nuclear hyperfine interactions can prolong the lifetime of certain collective many-body quantum states in coupled spin qubit arrays. The type of state preservation we discuss is qualitatively completely different from what is achieved by dynamical decoupling and other methods to combat decoherence (no designed external pulses are applied on the system). It is instead an intrinsic many-body entanglement effect in which the Overhauser disorder causes the state of the entire array to remain frozen despite the presence of nonzero interqubit couplings that would otherwise drive the spin system far away from its initial state. No control over the disorder is needed to observe this effect; only control over the exchange coupling between neighboring spins is necessary, and such control has already been demonstrated repeatedly in the laboratory. We consider spin qubit arrays in both GaAs and Si; although one would expect the effect to be more pronounced in GaAs due to the abundance of nuclear spins in this material, Si on the other hand allows the strength of the nuclear disorder to be tuned by changing the concentration of the spinful nuclear isotope Si$^{29}$. We propose in detail experiments to observe this effect and predict that disorder-induced state preservation can be seen in systems with as few as four spins, and therefore can be experimentally studied at the present time.

We consider an array of $L$ quantum dots containing one localized electron each subject to its own local magnetic field, $h_k$, coming from the Overhauser field generated by nuclear spins in the $k$th dot. If adjacent dots are tunnel coupled, there is an exchange coupling, $J_k$, between their electron spins, and the Hamiltonian is
\beq
H_{exchange}=\sum_{k=1}^LJ_k{\mathbf S}_k\cdot{\mathbf S}_{k+1}+\sum_{k=1}^Lh_kS_k^z.\label{ham}
\eeq
We include site-dependent exchange couplings $J_k$ because these couplings are generally not uniform in the real system. We note that Overhauser and charge noise affect $h_k$ and $J_k$ respectively. It is well established that the nuclear spins comprise a highly non-Markovian bath that evolves sufficiently slowly that it can be well approximated by a Gaussian quasistatic ensemble on timescales typical of electron spin dynamics, around a microsecond or less \cite{Cywinski_PRL09,Cywinski_PRB09,Cywinski_APPA11,Barnes_PRL12b,Martins_InPrep}. It is also known that charge noise has a non-Markovian character \cite{Medford_PRL12,Dial_PRL13}. We thus model both types of disorder by choosing the $h_k$ and $J_k$ from Gaussian ensembles with means $h_0$ and $J_0$ and standard deviations $\sigma_h$ and $\sigma_J$, respectively. In the case of charge noise, we truncate the distribution to the region $J_k>0$ since quantum dot exchange couplings are typically positive. The spin array is therefore modeled by a Heisenberg spin chain with random Gaussian on-site magnetic disorder and inter-site coupling disorder. Later on (see Eq.~\eqref{hamIsing} below), we also consider an Ising chain corresponding to an array of capacitively coupled ST or EO qubits.

We propose an experiment in which the array is initialized in a particular state and then allowed to evolve for a time $\tau$ that is large compared to the average interqubit coupling, $\tau\gg1/J_0$ (but small compared with the energy relaxation time $T_1$). At the end of the evolution, the probability that the array has returned to its initial state is measured. This experiment requires tunability of all the $J_k$ (but no tunability of $h_k$), although high accuracy is not necessary as we will see. If we take $L$ to be even, then we can view the spin chain as an array of $L/2$ ST qubits. The Hamiltonian of Eq.~\eqref{ham} conserves total $S^z=\sum_kS_k^z$; we focus on the $S^z=0$ subspace both because these states are easier to initialize and readout in the context of ST qubits and because the state-preservation effect is most dramatic in the largest subspace. A $S^z=0$ state can be prepared by first setting every other $J_k$ to zero to obtain a chain of uncoupled ST qubits, tilting each ST qubit into the (0,2) charge configuration to let it relax to a singlet state, and then adiabatically tilting back to the (1,1) configuration so that each qubit is brought to either the $\ket{\uparrow\downarrow}$ or $\ket{\downarrow\uparrow}$ state depending on which dot has the larger Overhauser field. All the $J_k$ should then be made comparable to each other prior to the start of the free-evolution period of the experiment. Other initial tensor product states can be prepared by additionally performing exchange pulses to swap pairs of up and down spins. Readout of the return probability is then performed by reversing these steps to map the final state onto the state in which each ST qubit is in a singlet. Each step in this process has already been achieved in laboratory spin qubit experiments.

Similar procedures can be used to initialize and readout an array of EO qubits in a $S^z=0$ tensor product state (now assuming that $L$ is a multiple of six), although an important difference is that these qubits live in the $S_{3spin}^z=1/2$ subspace of the three-spin Hilbert space. One approach is to initialize in a $S^z=0$ state by defining every other EO qubit to instead live in the $S_{3spin}^z=-1/2$ subspace and then initializing half of the EO qubits to $\ket{\uparrow\uparrow\downarrow}$ and the other half to $\ket{\downarrow\uparrow\downarrow}$. Alternatively, we could use states outside the $S^z=0$ space; the state-preservation effect should sill be visible although to a lesser extent. An array of LD qubits can be initialized to a $S^z=0$ state by first setting all the $J_k$ to zero, letting the system relax into the ferromagnetic ground state, and then applying $\pi$ pulses to half the spins.

\begin{figure}
\includegraphics[width=\columnwidth]{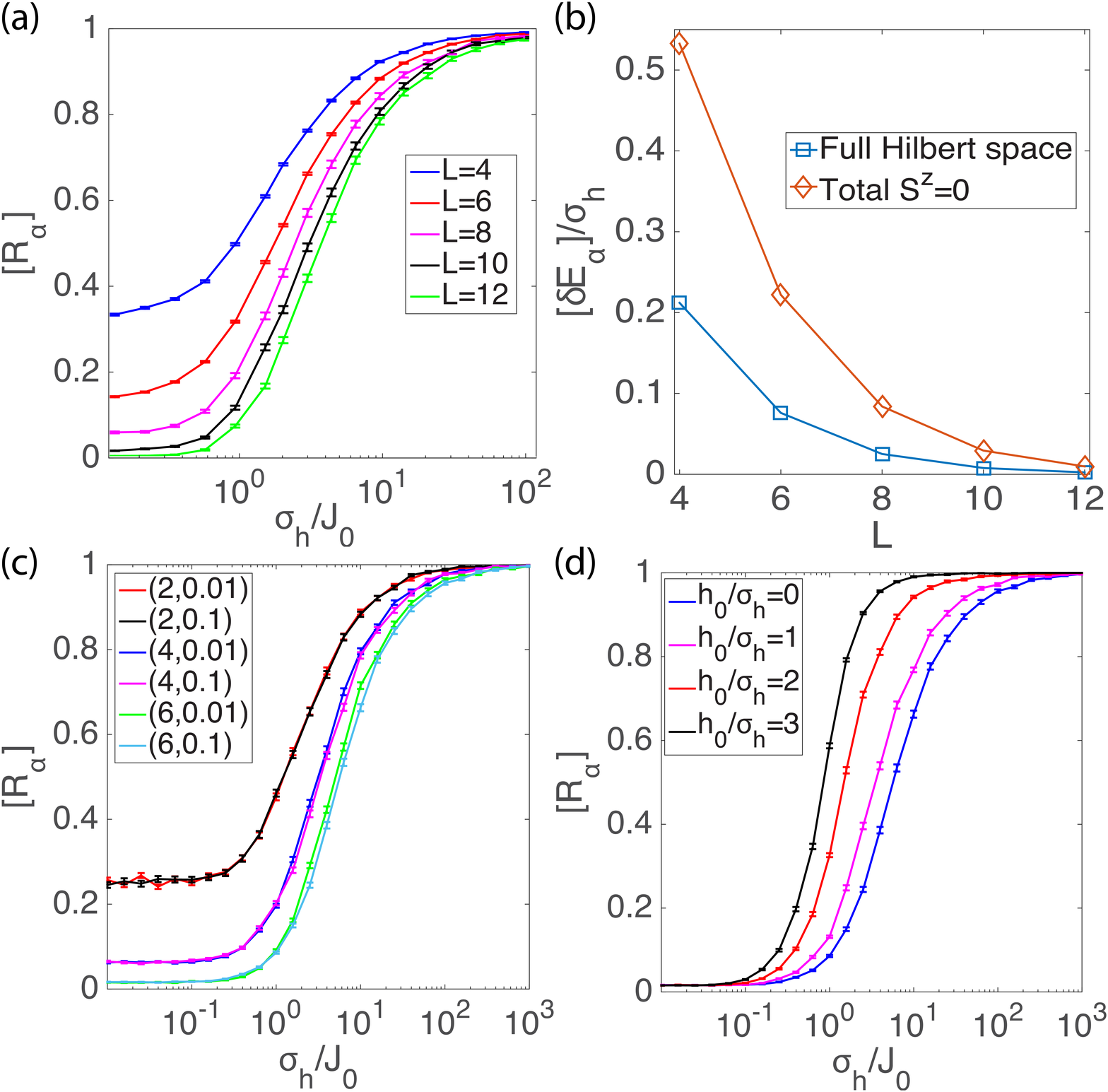}
\caption{\label{fig:RPforL4to12} State preservation from Overhauser noise. (a) Steady-state long-time return probability for $L=4...12$ exchange-coupled spins with $\sigma_J=0.1J_0$, $h_0=0$, $\tau=148.4/J_0$, averaged over $10^3-10^4$ disorder realizations. (b) Average non-interacting spectral gap versus system size. (c) Return probability for $L=2...6$ capacitively coupled ST qubits with $h_0=0$,  $\varepsilon=0.01,0.1$, $\sigma_J=0.1J_0$, and $\tau=148.4/J_0$. (d) Return probability for 6 capacitively coupled ST qubits with $\varepsilon=0.1$, $\sigma_J=0.1J_0$ for several values of $h_0$.}
\end{figure}

\begin{figure}[!t]
\includegraphics[width=\columnwidth]{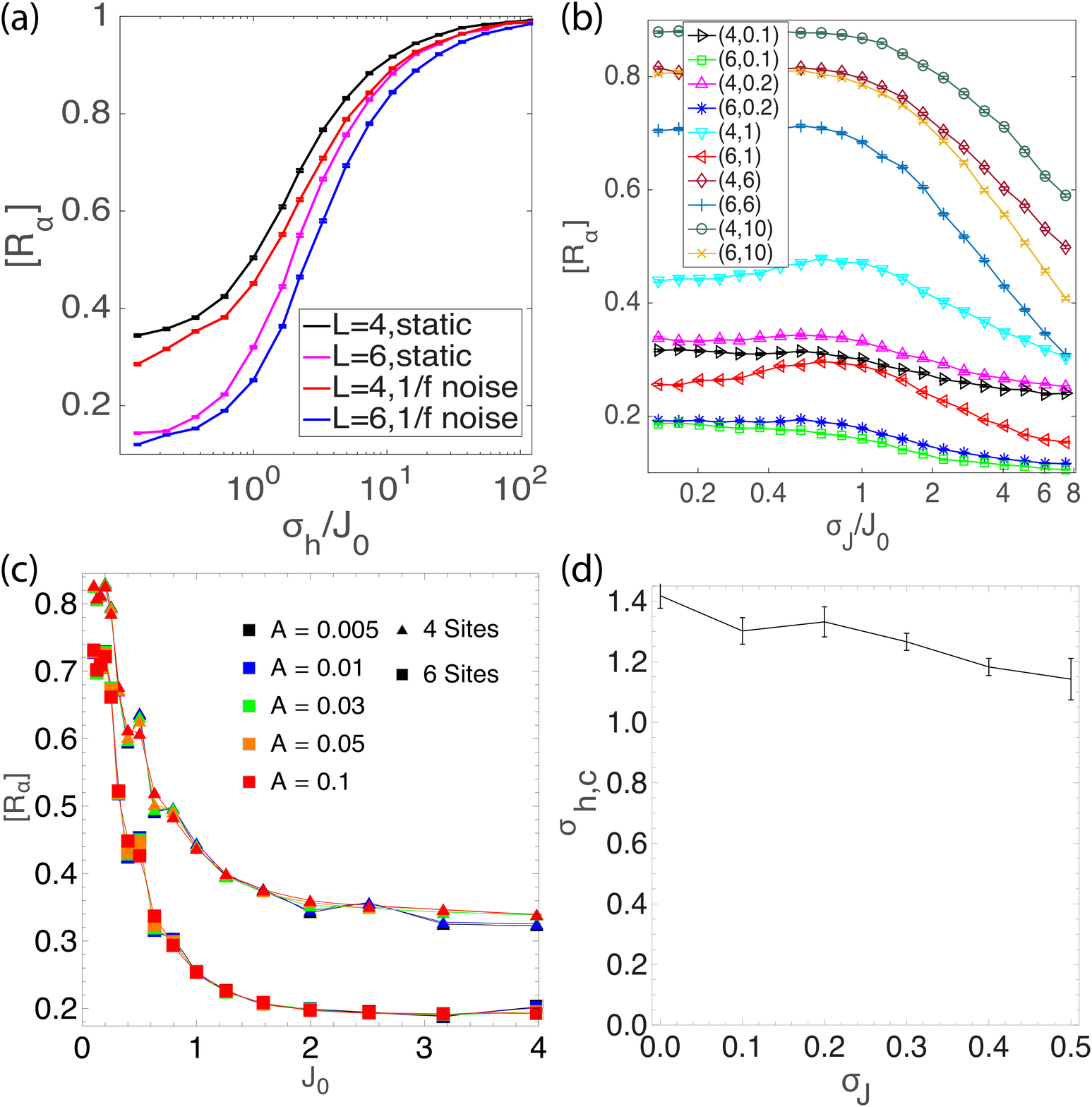}
\caption{\label{fig:RPvsSigmaJ} Effect of charge noise. (a) Comparison of return probabilities for static and $1/f$ charge noise. (b) Return probability as a function of $\sigma_J$ for $\sigma_h=0.1,0.2,1,6,10$ and $L=4,6$.  (c) Return probability as a function of $J_0$ with $\sigma_h=1$ for various values of $A$, where $A=\sigma_J/J_0$. (d) Critical $\sigma_h$ vs. $\sigma_J$.}
\end{figure}

Fig.~\ref{fig:RPforL4to12}a shows our exactly calculated prediction for the steady state (i.e., long time) return probability as a function of $\sigma_h/J_0$ computed from exact diagonalization and averaged over all $S^z=0$ initial states. We see that as the Overhauser disorder is increased beyond the average coupling ($\sigma_h\gtrsim J_0$), the likelihood of finding the array in its initial state after long times increases dramatically and quickly approaches 100\%, leading to the surprising conclusion that a sufficiently large amount of disorder {\it preserves} the state. Although this effect can be seen by varying the disorder for a fixed average spin-spin coupling as was done in generating Fig.~\ref{fig:RPforL4to12}a, it works just as well to vary $J_0$ while keeping $\sigma_h$ fixed since the behavior depends on the ratio of the two. This is an important point given that it is far more experimentally feasible to tune the couplings rather than the disorder in quantum dot arrays. We also note that although the global field $h_0$ is normally nonzero in experiments, we set $h_0=0$ in Fig.~\ref{fig:RPforL4to12}a for simplicity because it has no effect on the disorder-driven transition due to the conservation of $S^z$. 

One might guess that the enhancement of the return probability with increasing disorder could simply be due to the fact that typical energy spacings of the non-interacting spectrum scale linearly with $\sigma_h$ and thus become large compared to the exchange couplings for sufficiently large disorder, making the $J_k$ ineffective at mixing non-interacting eigenstates. To show that this is not a valid explanation for the noise-induced state-preservation effect, we show the average spectral gap size for the non-interacting Hamiltonian ($J_k=0$) as a function of system size in Fig.~\ref{fig:RPforL4to12}b. Even for $L=4$, the typical spectral gap remains smaller than $\sigma_h$ by a factor of two. This combined with the fact that Fig.~\ref{fig:RPforL4to12}a already shows a large return probability enhancement when $\sigma_h=2J_0$ reveals that a strong signature of state preservation is evident even when $J_k\sim\sigma_h$. For $L>4$, signatures of state preservation are apparent even when typical energy gaps are small compared to the $J_k$ due to the exponential suppression of energy gaps with system size. The state-preservation effect we predict is better understood as a remnant of many-body localization \cite{Nandkishore_ARCMP15}, although the latter exists only in the thermodynamic limit and is characterized by sharp transitions in quantities such as entanglement entropy or level statistics, whereas here we are predicting a few-body effect that occurs naturally in spin systems with realistic disorder and which is readily measured via the return probability. In the appendix, we support this interpretation by showing that quantities such as the entanglement entropy and level statistics behave consistently with other models believed to exhibit many-body localization. In the case of entanglement entropy, we show that the onset of state preservation is accompanied by a crossover from a volume to an area law.

In GaAs, $\sigma_h$ is typically on the order of tens of megahertz \cite{Foletti_NP09,Bluhm_PRL10,Martins_InPrep}, requiring $J_0$ to be in the megahertz range to achieve large return probabilities. In this case, the preservation effect will be visible for evolution times longer than a few microseconds. The transition from high to low probabilities could be probed either by increasing $J_0$ up to the hundreds of megahertz range or by reducing $\sigma_h$ via nuclear spin programming \cite{Foletti_NP09,Bluhm_PRL10}. In the case of silicon, the natural abundance of Si$^{29}$ nuclei produces a $\sigma_h$ on the order of one megahertz or less \cite{Wu_PNAS14,Kalra_PRX14}, while for isotopically-enriched silicon, the disorder is reduced much farther, down to the kilohertz scale \cite{Kalra_PRX14}. Thus, $J_0$ must be tuned orders of magnitude below the megahertz scale depending on the percentage of Si$^{29}$ and the evolution time must be increased accordingly; on the other hand, one could leave $J_0$ fixed and study the disorder-driven transition by using a set of samples with varying percentages of Si$^{29}$ so that $\sigma_h$ is varied directly.

Disorder-driven state preservation also happens in arrays of capacitively coupled spin qubits. In this case, we model each qubit as a pseudospin with two types of on-site disorder coming from Overhauser and charge noise within the qubit, and we approximate the capacitive coupling as an Ising interaction between pseudospins \cite{vanWeperen_PRL11,Shulman_Science12}:
\beq
H_{Ising}=\sum_{k<\ell}^L\varepsilon J_kJ_\ell\sigma_k^z\sigma_\ell^z+\sum_{k=1}^L(J_k\sigma_k^z+h_k\sigma_k^x).\label{hamIsing}
\eeq
Return probabilities for various array sizes and parameter values are shown in Figs.~\ref{fig:RPforL4to12}c and \ref{fig:RPforL4to12}d, where a clear transition is again visible in the vicinity of $\sigma_h/J_0\sim1$. Fig.~\ref{fig:RPforL4to12}c shows that this transition is not sensitive to the scale $\varepsilon$ of the capacitive coupling if $\sigma_h$ is varied while keeping $J_0$ fixed. On the other hand, Fig.~\ref{fig:RPforL4to12}d reveals that the transition does depend on the global magnetic field $h_0$ unlike the exchange-coupled case. In the case of the Ising interaction, total $\sigma^x$ (analogous to $S^z$ in the exchange-coupled case) is not conserved, and $h_0$ can have a substantial effect. The figure shows that by increasing the global magnetic field, the amount of disorder necessary for state preservation can be lowered significantly.

\begin{figure}[!h]
\includegraphics[width=\columnwidth]{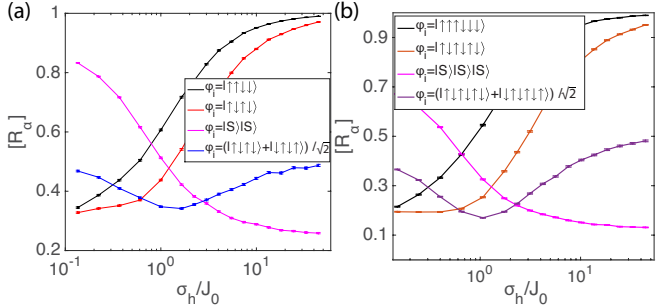}
\caption{\label{fig:RPvsState} Dependence on initial state. (a) Return probability as a function of $\sigma_h$ with $\sigma_J=0.1$ for 4 spins and for different initial states. (b) Similar to (a) but for 6 spins.}
\end{figure}

\begin{figure}[!t]
\includegraphics[width=\columnwidth]{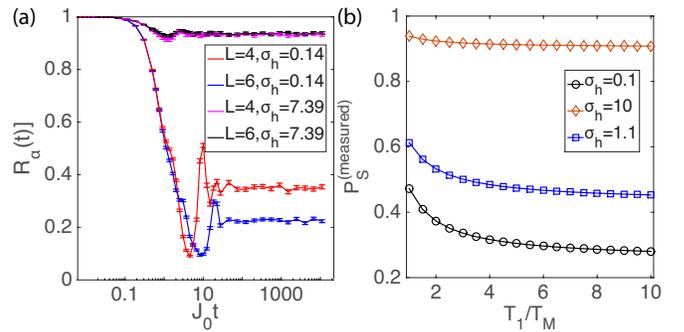}
\caption{\label{fig:RPvsTime} Dependence on time and relaxation. (a) Return probability as a function of time with $\sigma_J=0.1$ for 4 spins and for values of $\sigma_h$ above and below the critical value. (b) Singlet return probability as a function of relaxation time in units of the measurement time.}
\end{figure}

Charge noise is known to possess a nontrivial time-dependence characterized by a $1/f$ power spectrum \cite{Dial_PRL13,Medford_PRL12}. Fig.~\ref{fig:RPvsSigmaJ}a reveals that the state-preservation effect is insensitive to this time-dependence, at least for moderate levels of charge noise. To compare time-dependent and quasistatic charge noise, we equate the respective rms noise levels, which is equivalent to equating the integrated power spectral densities. Writing the $1/f$ power spectral density as $S(\omega)=A^2/\omega$, this is tantamount to choosing $A^2=2\pi\sigma_J^2/\ln(\omega_{uv}/\omega_{ir})$, where $\omega_{uv/ir}$ are frequency cutoffs. Here, we take $\omega_{uv}=100J_0$ and $\omega_{ir}=10^{-4}J_0$. The amount of charge noise (i.e., $\sigma_J$ or $A$) present in the system depends on the gate voltages and thus on the exchange couplings \cite{Hu_PRL06}. Typically, charge noise induces fluctuations in the exchange couplings on the order of a few percent or less, and thus when $J_0\sim\sigma_h$, we have $\sigma_J\ll\sigma_h$. The smallness of $\sigma_J$ relative to $\sigma_h$ ensures that charge noise has little effect on the Overhauser-induced state preservation (see Fig.~\ref{fig:RPvsSigmaJ}b). In fact, Fig.~\ref{fig:RPvsSigmaJ}b reveals that if charge noise becomes large enough, the return probability first increases before decreasing substantially, showing that intermediate levels of charge noise can actually enhance the state-preservation effect, while too much charge noise destroys the effect completely. This behavior stems from the fact that when $\sigma_J$ approaches $J_0$ there is an enhanced probability that a given $J_k$ will be close to zero, making it more difficult for the array to evolve away from its initial state. However, as $\sigma_J$ is made still larger, the probability that $J_k$ is larger than $J_0$ increases beyond 50\%, producing a more strongly coupled array and faster delocalization.

In actual spin qubit experiments the charge noise depends on the exchange coupling, and this dependence should be included if $J_0$ is varied since $\sigma_J$ will also vary. In Fig.~\ref{fig:RPvsSigmaJ}c, we consider a linear relation between $\sigma_J$ and $J_0$ which would follow from an exponential dependence of the exchange coupling on voltage fluctuations, a model that describes experiments well for low exchange couplings \cite{Dial_PRL13}. Fig.~\ref{fig:RPvsSigmaJ}c shows that there remains a strong signature of state-preservation even within this realistic charge noise model. Moreover, the critical value of $\sigma_h$, i.e., the location of the inflection point in the return probability as a function of $\sigma_h$, is nearly constant as $\sigma_J$ varies over a wide range (see Fig.~\ref{fig:RPvsSigmaJ}d), confirming that the state-preservation effect is insensitive to charge noise provided $\sigma_J/J_0$ is in the few-percent range typical of experiments.

So far, we have shown that Overhauser disorder preserves eigenstates of the non-interacting ($J_k=0$) Hamiltonian by computing the return probability and averaging over all tensor product states. However, it is not necessary to perform this average in order to see the state-preservation effect, a fact which greatly simplifies our proposed experiment. As shown in Fig.~\ref{fig:RPvsState}, an initial N\'eel-ordered state is preserved while a non-tensor product state such as a state made of pairwise singlets is not. Moreover, it can be seen in Fig.~\ref{fig:RPvsState} that different eigenstates are preserved to different degrees, i.e., there is a nontrivial initial state dependence. In particular, we see that N\'eel ordered states are less preserved compared to states with continuous strings of up and down spins. This is due to the fact that only adjacent up and down spins can be flipped by exchange interactions, so that states with few such adjacent pairs are naturally immune to local exchange interactions. In Fig.~\ref{fig:RPvsState}, we also consider what happens to the initial Schr\"odinger cat-like state $(\ket{\uparrow\downarrow\uparrow\downarrow}+\ket{\downarrow\uparrow\downarrow\uparrow})/\sqrt{2}$ and its six-spin analog. We find that, although as expected the return probability remains below 50\% regardless of the disorder, interestingly there is a non-monotonic dependence on the disorder in which the probability is reduced near the transition but then increases as the disorder is further increased.

Since the return probability is time-dependent, we investigate the behavior of the state preservation as a function of the final evolution time. Fig.~\ref{fig:RPvsTime}a shows that the return probability saturates on timescales large compared to $1/J_0$ and that for $\sigma_h\gg J_0$, the state never deviates far from its starting point. For a typical Overhauser disorder of $\sigma_h=30$MHz and taking $J_0\lesssim3$MHz to be well within the localization regime, we find that the return probability saturates at around $3\mu$s. This timescale is fast enough to justify the quasistatic nuclear noise model we are using and also slow enough that relaxation processes should not be important since these occur on the order of hundreds of microseconds or more for symmetric charge configurations \cite{Bluhm_NP11}. Decoherence and dephasing times are not relevant here since we are concerned with initial states that are energy eigenstates in the non-interacting limit.

Although relaxation in the symmetric charge configuration is negligible, relaxation during the measurement cycle (where it is dominated by unpolarized triplet to singlet relaxation) could have a more significant effect. The measurement cycle in ST and EO qubits is typically 5-10 $\mu$s, whereas $T_1\approx10-50\mu$s in the charge-imbalanced readout configuration. Since it is actually the singlet probability that is measured, this relaxation will lead to an artificial enhancement in the measured return probability \cite{Reilly_PRL08}, and it is important to quantify this to separate it from disorder effects. Fig.~\ref{fig:RPvsTime}b shows that relaxation is negligible in the strong disorder regime, but it can have a significant effect in the weak disorder region.

In conclusion, we have shown that nuclear spin Overhauser disorder can protect collective quantum states in spin qubit arrays containing as few as four spins. This is in stark contrast to its usual destructive role as a primary source of decoherence in these systems. The counterintuitive and rather spectacular effect we predict arises naturally in spin qubit arrays and can be measured using existing experimental capabilities already demonstrated in the laboratory. 

\acknowledgements
This work is supported by LPS-MPO-CMTC and IARPA-MQCO.

\appendix

\section{Appendix}

\begin{figure}
\includegraphics[width=\columnwidth]{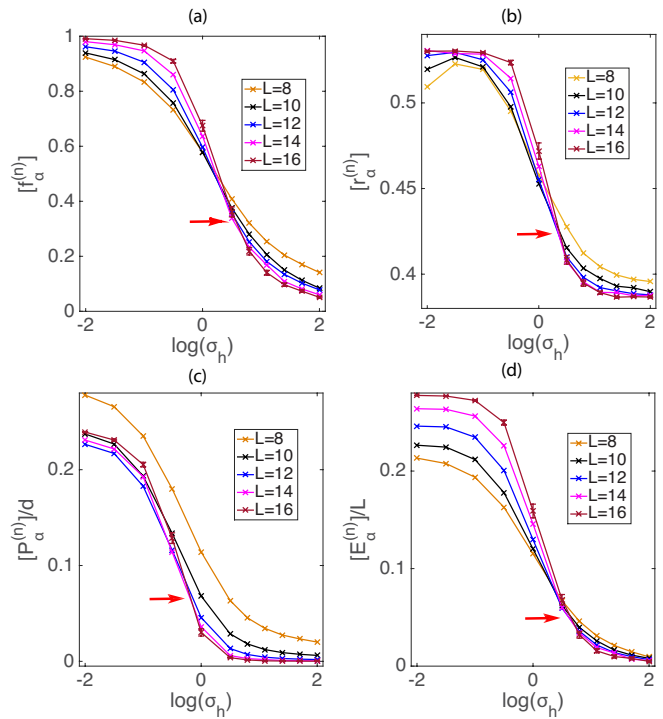}
\caption{\label{fig:MBLdiagnostics} Different quantities for characterizing many-body localized and delocalized
phases. (a) Dynamic polarization. In the localized region (large
$\sigma_{h}$) {[}$f_{\alpha}^{(n)}]$ decays to zero while in the
delocalized phase (small $\sigma_{h}$) the decay is small and this
distinction gets sharper as we increase the system size $L$. (b)
Level statistics. In the localized phase, the system exhibits Poisson statistics
($[r_{\alpha}^{(n)}]\approx0.39$) while in the delocalized phase
it exhibits Gaussian-orthogonal ensemble level statistics ($[r]\approx0.53$).
(c) Participation number divided by the Hilbert space dimension. In the
localized phase, $[P_{\alpha}^{(n)}]/d$ goes to zero as $L\rightarrow\infty$
while it remains finite for the delocalized phase. Here $d$ is the
Hilbert dimension. (d) Entanglement entropy divided by lattice size.
In the localized phase, $[E_{\alpha}^{(n)}]/L$ goes to zero as $L\rightarrow\infty$
(area law) while it remains finite for the delocalized phase (volume
law). For these plots, we have fixed $\sigma_{J}=0.1$, and the
number of random samples used are $10^{4}$ for $L=8,10,12$ and $60$
for $L=14,16$, respectively. The error bars are shown only for $L=16$.
The red arrow indicates the delocalized-localized transition point.
From these plots, we find the critical value of $\sigma_{h}^{c}=1.4\pm0.5$.}
\end{figure}

In this appendix, we provide evidence that the noise-induced state-preservation effect is a remnant of many-body localization that manifests as a localization-delocalization crossover in the regime of small system sizes. Here, we focus on the case of an exchange-coupled spin array, the Hamiltonian for which is given in Eq.~\eqref{ham}. The parameters $J_{k}$ and $h_{k}$ are independent random variables chosen
from normal distributions $f_{J}=\frac{1}{\sigma_{J}\sqrt{2\pi}}e^{-\frac{(x-J_{0})^{2}}{2\sigma_{J}^{2}}}$
and $f_{h}=\frac{1}{\sigma_{h}\sqrt{2\pi}}e^{-\frac{x^{2}}{2\sigma_{h}^{2}}}$,
respectively. We set $J_{0}=1$ as our unit energy; total $S^{z}=\sum_{k}S_{k}^{z}$
is conserved, and we focus only on the sector $S^{z}=0$ to simplify
the calculation. 

Here, we use the following different quantities to characterize the
many-body localization to delocalization transition:

\textbf{Participation Number (PN)}. This is defined as
\begin{eqnarray*}
P_{\alpha}(|\psi\rangle) & = & \left(\sum_{k=1}^{d}|\psi_{k}|^{4}\right)^{-1},
\end{eqnarray*}
where $d$ is the dimension of the Hilbert space and $\alpha$ denotes
a specific random realization. In the delocalized phase, $[P_{\alpha}]\propto d$
and in the localized phase, $[P_{\alpha}]$ is roughly independent
of system size $L$. Throughout this appendix, $[\;]$ denotes an average over sites,
states and samples. 

\textbf{Dynamic Polarization}.
We can consider the evolution of an initially prepared inhomogeneous spin density mode with maximal wavelength, $\mathcal{M}=\sum_{k}S_{k}^{z}\exp(i2\pi k/L)$.
We then define the dynamic polarization as \cite{Pal_PRB10}:
\begin{eqnarray*}
f_{\alpha}^{(n)} & = & 1-\frac{\langle n|\mathcal{M}^{\dagger}|n\rangle\langle n|\mathcal{M}|n\rangle}{\langle n|\mathcal{M}^{\dagger}\mathcal{M}|n\rangle}.
\end{eqnarray*}
We expect that $[f_{\alpha}^{(n)}]\rightarrow1$ in the delocalized phase and $[f_{\alpha}^{(n)}]\rightarrow0$
in the localized phase \cite{Pal_PRB10}.

\textbf{Level statistics}. We denote the level spacings by $\delta_{\alpha}^{(n)}=|E_{\alpha}^{(n)}-E_{\alpha}^{(n+1)}|$. The ratio of adjacent gaps is then 
\begin{eqnarray*}
r_{\alpha}^{(n)} = \min\{\delta_{\alpha}^{(n)},\delta_{\alpha}^{(n+1)}\}/\max\{\delta_{\alpha}^{(n)},\delta_{\alpha}^{(n+1)}\}.
\end{eqnarray*}
In the absence of charge noise, $[r_{\alpha}^{(n)}]\approx0.53$ in the delocalized phase with Gaussian-orthogonal
ensemble level statistics, while $[r_{\alpha}^{(n)}]\approx0.39$ in
the localized phase with Poisson level statistics \cite{Pal_PRB10,Luitz_PRB15}.

\textbf{Entanglement entropy}. For an eigenstate $|n\rangle$, we do a partial trace to obtain the
reduced density matrix for the subsystem A: $\rho_{A}=\text{Tr}_{B}(|n\rangle\langle n|)$.
The entanglement entropy is defined as 
\begin{eqnarray*}
E_{\alpha}^{(n)} & = & -\rho_{A}\log\rho_{A}.
\end{eqnarray*}
We expect that in the delocalized (localized) region, the entanglement entropy obeys a
volume (area) law \cite{Luitz_PRB15}.

Fig.~\ref{fig:MBLdiagnostics} shows our results for each of these quantities. In all cases, the results are fully consistent with the existence of a many-body localization phase transition in the thermodynamic limit, $L\to\infty$. We estimate the critical noise level to be $\sigma_{h}^{c}=1.4\pm0.5$.

\end{document}